# Assessment of Sign Language-Based versus Touch-Based Input for Deaf Users Interacting with Intelligent Personal Assistants

Sign Language-Based versus Touch-Based Input for Deaf Users with IPAs

Accessibility, advantages, and disadvantages of sign language interaction with intelligent personal assistants

NINA TRAN*

Gallaudet University, nhan.tran@gallaudet.edu

PAIGE S DEVRIES

Gallaudet University, paige.devries@gallaudet.edu

MATTHEW SEITA

Gallaudet University, matthew.seita@gallaudet.edu

RAJA KUSHALNAGAR

Gallaudet University, raja.kushalnagar@gallaudet.edu

ABRAHAM GLASSER

Gallaudet University, abraham.glasser@gallaudet.edu

CHRISTIAN VOGLER

Gallaudet University, christian.vogler@gallaudet.edu

With the recent advancements in intelligent personal assistants (IPAs), their popularity is rapidly increasing when it comes to utilizing Automatic Speech Recognition within households. In this study, we used a Wizard-of-Oz methodology to evaluate and compare the usability of American Sign Language (ASL), Tap to Alexa, and smart home apps among 23 deaf participants within a limited-domain smart home environment. Results indicate a slight usability preference for ASL. Linguistic analysis of the participants' signing reveals a diverse range of expressions and vocabulary as they interacted with IPAs in the context of a restricted-domain application. On average, deaf participants exhibited a vocabulary of 47 +/- 17 signs with an additional 10 +/- 7 fingerspelled words, for a total of 246 different signs and 93 different fingerspelled words across all participants. We discuss the implications for the design of limited-vocabulary applications as a stepping-stone toward general-purpose ASL recognition in the future.

---

* Nina Tran and Paige DeVries contributed equally to this paper.

## 1 INTRODUCTION

Recent developments in intelligent personal assistant (IPA) technology show that these smart interfaces are rapidly growing in popularity, for home and family use [16][2]. Amazon Alexa [39], Apple Siri [5], Google Home Assistant [29] are examples of voice-activated interfaces that communicate with users through their respective systems using Automatic Speech Recognition (ASR), which translates human speech into text [22]. Studies have indicated that deaf and hard of hearing users, as well as those who have speech-related disabilities, are interested in interaction with IPAs [24][19][9][46]. These systems often have designated activation phrases, such as the wake commands "Alexa" or "OK Google" [39], and their functionality generally depends on responding to voice commands. Intrinsically, voice-controlled IPAs are inaccessible to deaf users, even those who use their voice frequently to communicate [24][25].

The main challenge faced by researchers in this realm of human-computer interaction (HCI) is how deaf users can interact with IPAs [48][36]. IPAs and other voice-controlled smart interfaces heavily depend on audio input. Text-to-speech has limited reliability and is not a functionally equivalent experience for deaf users. Few IPAs offer text input, but only on specific, compatible devices [30], and while this is also not functionally equivalent, it is often not a desired, natural, nor efficient method of input. Current state-of-the-art ASR continues to struggle with "deaf speech," being unable to achieve consistent results. Additionally, preliminary testing indicates that any form of text input is significantly slower—3 to 4 times—compared to speech, which interferes with usability [22]. IPAs are currently incapable of recognizing sign language commands, as highlighted in prior research [27][13][36][20][37].

IPAs also prioritize audio output modalities, especially with devices that do not have a display screen (e.g., smart speakers). While technological developments have addressed this lack of output accessibility by adding smart displays with captioning for the audio output, it is not always available. Additionally, this addresses the output side of the device interaction, which would not work without input. There is a scarcity of research focused on the understanding of the deaf user experience of interacting with IPAs, particularly regarding aspects related to accessibility [48].

Recent research suggests that researchers may find success in utilizing the Wizard-of-Oz method to better understand how deaf users respond to different user interface designs [13]. Using this methodology, we can simulate machine sign language recognition technologies to better gauge their usability. In this paper, we address IPA accessibility via American Sign Language (ASL) from an interdisciplinary standpoint spanning accessibility, HCI, interpretation, and linguistics. Our specific purpose is to tackle two calls to action as referenced in [13]: "Call 2: Focus on real-world applications" and "Call 3: Develop user-interface guidelines for sign language systems." This study provides an empirical analysis of how deaf users prefer to interact with/'] these systems, addressing these three research questions:

- RQ1: What are the perceptions and preferences of deaf users regarding sign language-based versus touchscreen-based input methods while interacting with intelligent personal assistants (IPAs)?
- RQ2: What is the nature of the interaction between deaf users and IPAs within the confines of a limited-domain application?
- RQ3: What is the range of expression and vocabulary seen among deaf users interacting with IPAs within the confines of a limited-domain application?

## 2 BACKGROUND AND RELATED WORK

For those who are deaf, hard of hearing, or someone with hearing loss, sign language is an important aspect of communication. There are over 200 known sign languages around the world, and around 70 million deaf people use sign language as their primary mode of communication in their daily lives [53]. Within the United States, there are approximately 37.5 million people reporting hearing loss [44], and among them, nearly 500,000 people primarily communicate with American Sign Language (ASL) [43]. This constitutes a substantial sub-population facing limited access to IPAs due to the technology's reliance on verbal communication modalities.

### 2.1 Automatic Speech Recognition

ASR enables IPAs to understand spoken commands by using technology to translate spoken language into written text. Despite the quick development of ASR technologies [13][26][41], studies have revealed that ASR yields substantially higher variability with deaf user's speech compared to the data that was used to train ASR models [12][22][24][31]. People with dysarthria encounter similar challenges while interacting with IPAs [9], and it is more common for children or older adults to experience dysarthric speech [54]. These speech-related disabilities often result in speech that is unintelligible and beyond state-of-the-art ASR capabilities.

State-of-the-art ASR engines typically achieve around 5-6% Word Error Rate (WER) for people with normal speech but is significantly higher for people with deaf speech and speech-related disabilities, demonstrating that ASR technology currently lacks the capability to accurately decipher deaf speech [25]. ASR technology, the foundation that IPAs are built upon, requires auditory input and remains inaccessible, especially for those who are deaf and rely on sign language for communication [27][48][52].

### 2.2 Input methods and Deaf Interest in IPAs

Depending on how they interact with a system, deaf users have diverse needs and preferences in varying contexts. Turk [51] states that having many modalities available to interact with a system attracts users who might otherwise not use said system due to disabilities. This increased access could improve efficiency and help people become more adaptable and autonomous in continuously changing environments and situations. For example, a deaf person may ask their IPA to turn on the lights at a certain time in the morning, so they get up on time for work.

Some augmented and alternative communication (AAC) solutions can help address some of the deaf speech and speech-related issues, including mobile apps (e.g., Apple's VoiceOver, VoiceNavigator [17], DiscoverCal [23]), wall-mounted touchscreens, and voice-control speakers (e.g., Amazon Echo, Apple Siri, and Google Assistant) [40]. In contrast to voice control, all-mounted touchscreens, and apps, Luria et al. [40] found that participants who used embodied social robots had the highest situation awareness -- embodied social robots provide visual information on what they are doing and enable participants to think while they interact rather than relying only on vocal commands.

Prior studies have shown that deaf users are less familiar with IPAs than the general population. In a study by Lopatovska et al., it was found that although 52% of deaf participants were satisfied with Alexa, 42% of them were extremely dissatisfied, and that users often expressed more comfort in using Alexa as a tool for low-risk requests such as checking the weather or cueing music [39]. Researchers continue to explore alternative input methods, including ASL, Text-to-Speech (TTS), gestures, and deaf speech. Most have found that deaf participants prefer to use ASL as an input method when interacting with IPAs [12][24][26][27][41][48][50][52][31]. There have been calls for research to consider the accuracy with which wake-up interactions can be recognized (analogous to speaking "Alexa" or "Hey Google") [41].

Studies that focus on ASL for IPA interaction often employ a Wizard-of-Oz method where a hearing person (I.e., a trained ASL interpreter) in a different room performs the ASL-to-Speech function simulating machine understanding [48][52]. This method has also been applied to gestural systems [48]. There are some limitations inherent in ASL-based Wizard-of-Oz approaches, such as time delays between ASL input and device response, and a lack of live transcription of the Alexa commands [52]. Other limitations pertain to a system not recognizing custom “home signs,” and that users did not know the list of signs a system could recognize [48].

### 2.3 Sign Language Recognition

In the past, experts have pursued the development of sign language processing within the confines of their specific fields of expertise, often lacking substantial collaboration with related fields of study. An ideal, comprehensive sign language recognition system would involve a multidisciplinary team comprising experts and advisors with substantial experience in computer vision, computer graphics, natural language processing, human-computer interaction, linguistics, and deaf culture [13].

Prior research has suggested recommendations for improving sign language recognition technologies, emphasizing the significance of focusing on real-world applications and developing user-interface guidelines for sign language systems [13][21][19]. In contrast to the development of ASR, automatic sign language recognition is much less developed due to several factors: (1) continuous sign language recognition is required when translating sign language in real-world scenarios; (2) sign languages do not have the written form since the majority of machine translation and natural language processing relies on written languages; (3) annotations require a substantial amount of time (i.e., one hour of annotation per one minute of ASL) and are prone to human errors, therefore proficient annotation experts must undergo extensive training [13][19].

ASL has several unique phenomena distinct to signed languages, including classifiers, fingerspelling, non-manual signs, depiction, and role-shifting. Any developing sign language recognition system needs to be trained on these phenomena and be able to recognize when a user is intending to use them. These significant linguistic features (in combination with other cultural and societal features) contribute to a wide range of fluency in sign languages that requires processing technology to have a complex understanding of sign language interaction [13].

Since the datasets used in developing algorithms to train technology to recognize sign language are often nonrepresentative of real-world uses, interdisciplinary collaboration would allow professionals to combine their datasets (which each focus on their specific field) into a more comprehensive and applicable algorithm for deaf users interacting with IPAs. The existing ASL datasets are constrained to alphabet fingerspelling and isolated signs, such as those developed by ASL Citizen, which has compiled a sizable corpus of sign language (2.7k unique ASL signs as of the date) [1].

Additionally, Microsoft has developed an algorithm to combat the “Midas Touch” problem with gestural systems where any gesture awakens the device, by combining facial features, body pose, and motion to determine when the device is intentionally being activated [50], which proves that technology is capable of being taught to recognize gestures. Using gestural systems may be advantageous for those with speech-related difficulties since they are more interactive.

## 3 RESEARCH METHODS

We designed a within-subjects repeated measures study to compare using ASL, smart home apps (“Apps with Alexa”) and Tap-to-Alexa as interaction methods with an IPA in a limited-domain smart home environment, using an Amazon Echo Show device. Due to the Echo Show’s inability to support ASL, this experiment employed a Wizard-of-Oz approach to issue sign language commands to Alexa. In this section, we describe the participants, materials, study design and methods in more depth.

### 3.1 Recruitment and Participant Demographics

We recruited a total of 23 deaf participants for an in-person study. The eligibility criteria for participants were: (1) identifying as deaf or hard of hearing; (2) fluency in American Sign Language, and (3) being at least 18 years old. Participants were asked to complete an intake survey with demographic information and prior experience with IPAs.

Deaf participants identified themselves as follows: 13 as female, 8 as male, and 2 as non-binary. 19 self-identified as deaf, 3 as hard of hearing, 1 as deafblind. Most (65%) self-identified as White or Caucasian, while the remaining participants (13%) identified as Hispanic or Latino, 9% as African American or Black, 9% as Asian, and 4% as Multiracial. Participants were on the younger side, with 52% being between 18-24, 7% between 25-34, 13% between 35-44, and 4% between 45-54 years old. All participants use ASL regularly while interacting with others in person, some use also written English (52%), spoken English (39%), and signed English (21%) in addition to that. All except two participants use ASL as their primary mode of communication, with the other two using spoken English. Their educational backgrounds ranged from high school diploma or GED (30%) to some college or no degree (22%), associate degree (9%), bachelor's degree (30%), and graduate or professional degree (17%).

Participants were asked to rate their level of expertise with smart technology (e.g., smartphones and intelligent personal assistants). Most reported proficient experience (52%) with the remainder reporting advanced experience (30%) and some experience (17%). Most participants (52%) rarely use voice control interfaces, while others (35%) never use them at all. Only 9% use them more than 3 times a week, and 4% use them 1-2 times a week. Most (74%) do not own a smart home control device. When asked whether they could imagine owning one, most (47%) indicated maybe, followed by yes (35%), while the remainder (17%) were undecided; no one answered no. The majority reported that family members or close friends (57%) own a smart home control device, and others reported someone they live with (22%) or someone they visit often (22%) has one.

### 3.2 Materials

Here we describe the hardware and equipment used, usability instruments, post-experiment surveys, and task lists provided to participants for interacting with the limited-domain smart home environment.

#### *3.2.1 Equipment*

The setup consisted of two different stations, the "Dorothy" station, and the "Wizard" station. The former was set up for participant interaction, while the latter supported the hidden researcher to support the Wizard-of-Oz approach from a separate room. In particular, the "Dorothy" station provided the smart home environment to participants, centered around an Amazon Echo Show device. It was set up to provide responses in both spoken English on-screen captioning. Additionally, there were two Philips Hue [45] multi-color lights, a Fire TV, two video cameras, and an EarFun UBOOM 28W speaker aimed at the Echo Show. The two video cameras were set up across from each other to capture both what the participant was signing (front camera) and the responses they received from the smart home environment around them (back camera). There also was a laptop driving the front camera. It was controlled remotely by the researcher at the "Wizard" station.

At the "Wizard" station there were a laptop, additional monitor, and Blue Yeti microphone all equipped to aid in the interpreting of the ASL to Alexa. To maintain the audiovisual connection to the "Dorothy" station, the laptop was running Facetime [7] to connect to the front camera and see the participants' signing, the Photo Booth app [6] used to record the interaction for analysis, and VNC Viewer [47] to control the laptop in the "Dorothy" room. A more detailed description of the set-up is provided in Appendix A.5.

*3.2.2 In-Experiment Instruments and Post-Experiment Survey*

The System Usability Scale (SUS) is a foundation for assessing the usability of a system [14][49]. It uses a 5-point scale ("Strongly Agree" to "Strongly Disagree") across ten questions, which is then scaled and normalized to a 100-point score. However, the SUS is administered in written form, which raises validity concerns for some deaf signers for whom English is not their first language. In a worst-case scenario, participants may be excluded from usability studies, as these instruments do not take into account the use of sign languages and the language barriers the scale operates with. To counteract this, we gave participants a choice between using the written English SUS version and/or the ASL-SUS developed by Berke et al [11]. ASL-SUS emphasizes dynamic equivalence between ASL translations of the survey items and the original English text, rather than a word-for-word translation that may not accurately embody the concepts discussed [32]. The ASL-SUS has been psychometrically validated as giving equivalent results to the SUS [11]. Participants completed the SUS for the three different conditions in our experiment (ASL, Tap-to-Alexa, and Apps with Alexa).

In the post-experiment survey, deaf participants were asked how they felt about navigating these systems (e.g., fingerspelling vs typing movie titles to Alexa) and their preferred wake-up methods (analogous to speaking the word "Alexa" to activate the IPA). We also asked deaf participants their preferred input methods and whether they could imagine using Alexa to control smart home devices. Additionally, we asked deaf participants to rate the importance of various features of ideal smart home systems.

*3.2.3 Task Lists*

To compare the input methods, we asked the deaf participants to complete three parallel task lists (named A, B and C; see Appendix A). We counterbalanced both the task lists and the sequence of conditions for each participant to mitigate biases related to the order of tasks. This approach was adopted to prevent any potential biases induced by differences across the task lists or order of conditions. Participants were given each of the three work lists that corresponded to the three input methods (conditions), which we rotated as a facet of counterbalancing. For instance, the first participant started with A, B, and C, followed by the next participant beginning with A, C, and B. This rotation continued for subsequent participants to ensure an unbiased distribution across participants. The average session length is 35 minutes.

The task lists each featured a mix of action items for interacting with devices, selecting videos for playback, and setting timers. Participants were asked to:

- turn on/off the Fire TV
- turn on/off lights, change the color of lights, and change the brightness of lights
- select movies to watch on the Fire TV, fast-forwarding, rewinding, pausing, resuming, and ending
- setting and interacting with timers

The selected tasks aimed to mirror interactions that a person might have with an IPA in a standard smart home environment, with a focus on limiting the domain. To ensure that even participants new to IPAs would know what to do, the tasks were presented in plain English, and each task list consisted of a specific set of Alexa commands. Furthermore, the selection of Alexa commands was informed by diverse signs in ASL, providing an opportunity for a deeper understanding of participants' interactions with an IPA. For instance, the signs for "Fire" and "turn on/off" exhibit several variants, illustrating the rationale behind selecting these specific tasks. After going through the task lists, participants typically became more confident. We provided them with a five-minute window for free play following the task completion (which was done before completion of the SUS). During this time, they frequently utilized the time to check the weather and browse YouTube.

## 3.3 Methods

### *3.3.1 Participant Procedures*

Each participant experiment session was up to one hour. In the beginning, deaf participants were asked to enter the simulated smart home environment. Researchers who were fluent in both ASL and English were to provide any necessary clarification about the informed consent process and procedures in the participant's preferred language(s). We asked participants to fill out an intake survey with their demographic information and familiarity with IPAs and related technologies. They were informed about the purpose of the study and what they could expect from the following in-person portion of the study, including a basic introduction to the devices and the SUS. After the intake survey was completed, deaf participants executed tasks while interacting with Alexa across the three conditions: ASL to Alexa, Tap to Alexa and Apps with Alexa.

For **ASL to Alexa,** using the Wizard-of-Oz method, deaf participants interacted with Alexa by signing in ASL towards the Amazon Echo Show, following the task list. They remained unaware of the "Wizard" human ASL interpreter that was verbally communicating the sign language commands to Alexa. Figure 1 illustrates a study participant signing an ASL command to Alexa. Participants were instructed to sign into the front camera (perched atop the device), which, to their understanding, was responsible for capturing their signing. They were also made aware that their signing would be videotaped from the front and back, which would subsequently be analyzed by researchers on our team. Figure 2 depicts the Wizard live-translating a participant's ASL command. From the participants' perspective, it seemed as if Alexa understood the signed commands and responded via captions on the Echo Show screen. Note that participants did not see captions for the spoken commands issued by the Wizard.

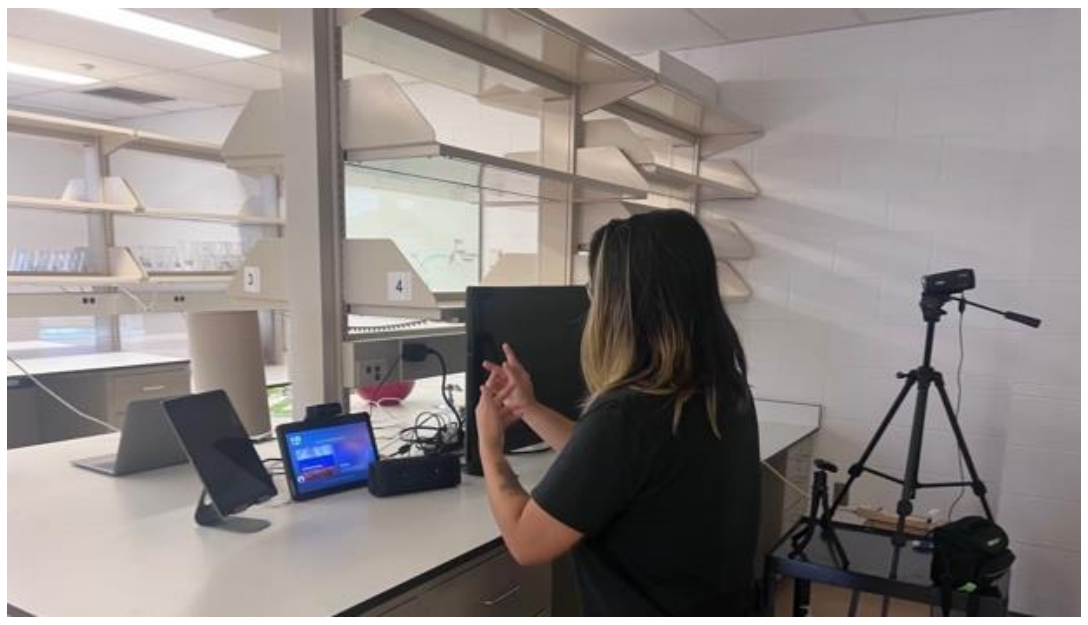

Figure 1: A participant signs a command to Alexa while on camera via a FaceTime link.

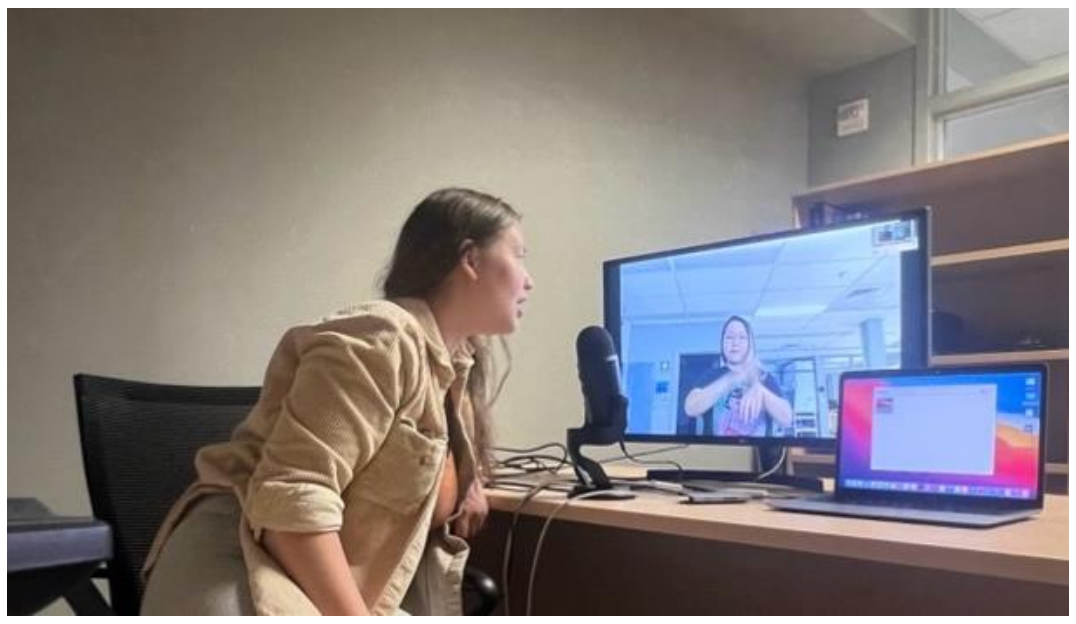

Figure 2: The interpreter behind the scenes (a.k.a. Wizard) translating a command to Alexa in a separate room, while viewing the participant's signing on a computer monitor linked via FaceTime.

For **Apps with Alexa,** participants were instructed to use the iPad to follow the corresponding task list and interact with the smart home environment. They used a mix of the Philips Hue app to control the lights, the Alexa app to set timers [4], and the Fire TV app to control the TV and video playback. Some participants also opted to use the Fire TV remote in place of the app. Researchers stayed at hand to help participants with using the apps.

For **Tap to Alexa,** participants were instructed to interact with the Echo Show device's touchscreen [3]. They scrolled through a list of labeled icons representing commands and picked the appropriate ones for the task at hand. All icons had been preconfigured by the research team, so the main challenge in this task was scrolling and identifying the correct one. As in the ASL to Alexa condition, Alexa's responses were shown in captioned form on the Echo Show.

After completing each condition and engaging in free play with Alexa, the participant completed the SUS. After completing all conditions and tasks, the participant filled out the post-experiment survey.

### *3.3.2 Wizard Procedures*

To facilitate communication between deaf participants and Alexa in the ASL to Alexa condition in a manner representative of potential future ASL recognition systems, the Wizard was required to use literal interpretation for each session. This approach is in marked contrast to what interpreter training typically practices, as most other scenarios require interpreters to use functionally/dynamically equivalent interpretation. In the latter situation, interpreters voice the concepts that are implied by what they see [42]. This created tension for the Wizard and required them to be continually monitored by the deaf research team to ensure they did not inadvertently step out of their role. To further minimize possible interpretation biases, the Wizard was only informed of the condition order, but not of the specifics of each task list.

The Wizard specifically looked for the wake word. Participants were allowed to either fingerspell Alexa or use a name sign for Alexa (e.g., “FS(ALEXA)” or “NS(AX);” see also Section 3.3.3). If participants omitted the wake word, the Wizard did not speak Alexa’s name, but still interpreted the command produced by the participant. Note that some participants encountered difficulty in signing the wake word and recalling it due to their limited exposure to IPA technology (but see also the discussion of eye gaze and hand waving in Section 5.2).

If Alexa did not receive a command, it was up to the participant to become aware. If there was no response from Alexa, the Wizard did not repeat the command and instead waited for the participant to decide how to proceed with the task. Some participants chose to repeat the task item, others chose to move on to the next one. Note that the Echo Show has a visual indicator showing whether the wake word has been uttered or when a command is being processed in the form of a blue line at the bottom of the display screen, and its presence or absence could provide clues to the participants. If the Wizard failed to understand the participant’s signing, they spoke the command “Alexa, write,” which forced the response “I’m sorry, I didn’t get that” via audio and captions on the Echo Show. This typically prompted the participant to try their command again.

### *3.3.3 Data Analysis*

We performed both descriptive and inferential statistical analyses on the SUS scores via paired t-tests with Bonferroni correction for each of the three conditions. We also performed an Analysis of Variance (ANOVA), to compare the group means across the conditions. Additionally, we calculated descriptive statistics for the post-experiment survey questions.

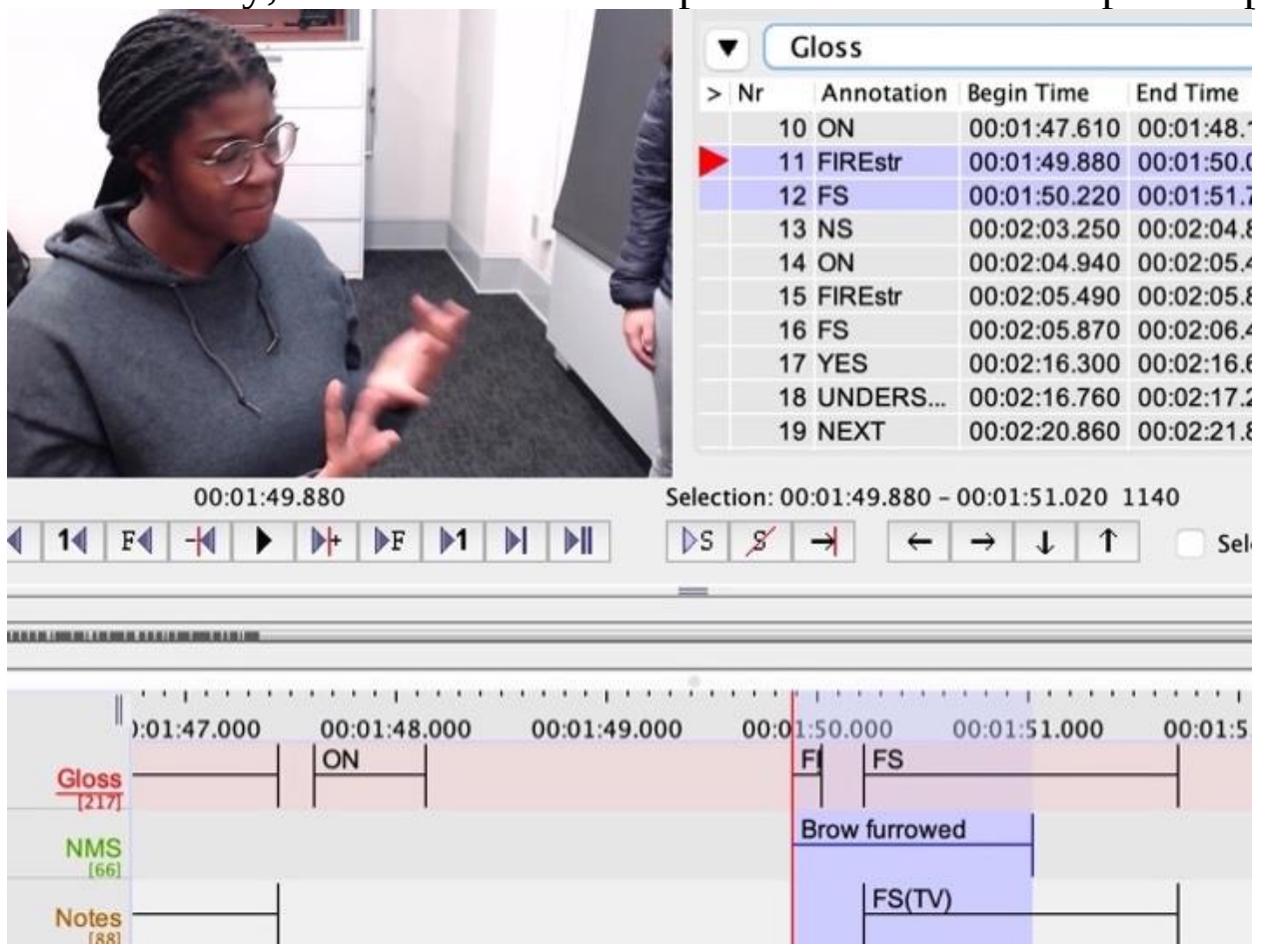

Figure 3: Snapshot of ASL annotations in the gloss, non-manual signals (NMS) and notes tier. The participant is performing the sign for “FIRE” with the straight movement variant, and has their brows furrowed at the same time. The next sign is fingerspelled, with the notes tier showing that the word “TV” is being fingerspelled.

Furthermore, we analyzed the ASL usage of deaf participants via annotating their videos in the ELAN linguistic annotation platform [18], and subsequently calculating statistics on usage, vocabulary size, as well as screening the annotations for linguistic phenomena that may affect the design of a future automatic sign language recognition system for IPAs. Although sign language annotations can go into minute details on hand configuration, body movements, and each part of the face, for this study we focused on the most salient characteristics needed to interact with an IPA. To this end, we annotated three tiers of information: gloss, non-manual signals (NMS), and notes, as shown in Figure 3.

Our annotation process followed the conventions used by the ASL Signbank, which is “a collection of ASL signs linked with ID glosses, meant for use by fluent ASL signers as an annotation tool for ASL videos with ELAN and the ASL SLAASh [Sign Language Annotation, Archiving and Sharing] conventions” [34][35]. Following these conventions ensures machine-readability through searchable and countable data and ensures that signs are represented consistently across datasets and annotators.

The gloss tier provides a representation of each respective sign in English though ID glosses. These provide a standardized way to link the observed signs (and any dialect-induced variants) to items in the Signbank database, which consist of ID glosses, translations, and ASL video entries [33]. Note that the English representation in the ID gloss is not necessarily a translation of the sign in the context in which it occurred, but it does uniquely identify the sign and the form in which it was presented. The NMS tier was used for notating significant participant reactions within their interactions while signing ASL to Alexa. The notes tier is necessary, because transcription using glosses does not cover all ASL linguistic phenomena and includes some codes that require further annotation explaining the specific occurrence, for example, fingerspelling, name signs, and gestures. More specifically, if fingerspelling occurred, the notes tier contains the exact fingerspelled word; if a name sign occurred, the notes tier contains the name of the referenced entity; and if a gesture occurred, the notes tier contains further information on the characteristics of the observed gesture.

## 4 RESULTS

In the following, we provide the usability results, participant preferences from the post-experiment survey, and findings regarding ASL usage with IPAs.

### 4.1 Usability Results

Figure 4 shows the mean SUS for the ASL, Tap and App conditions. ASL was preferred, with a mean SUS of 71.6 (SD 16.428, SE 3.427). Tap to Alexa had the next-best usability score, with a mean SUS of 61.4 (SD 19.304, SE 4.025). Apps ranked last with a mean SUS of 56.3 (SD 26.670, SE 5.561). The one-way repeated-measures ANOVA for the SUS showed statistical significance with $p = 0.015$. The results of the post-hoc paired t-testing with the Bonferroni correction is shown in Table 1. None of the pairwise differences between ASL, Tap and Apps rose to the level of statistical significance, although ASL vs Apps came close with $p = 0.051$.

Table 1: Paired t-test results on the SUS scores across the three conditions.

| Conditions | t-statistic | df | p |
|---|---|---|---|
| ASL vs Apps | -2.58 | 22 | 0.051 |
| ASL vs Tap to Alexa | 2.47 | 22 | 0.065 |
| Apps vs Tap to Alexa | -0.999 | 22 | 0.987 |

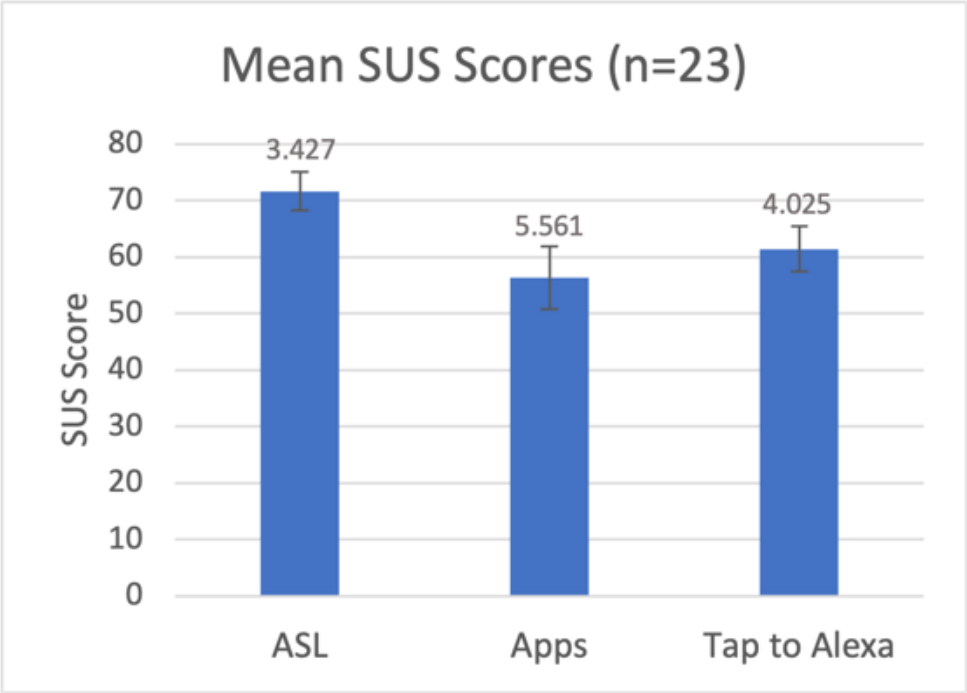

Figure 4: A bar graph showing the mean SUS scores for each of the three conditions with Alexa. None of the pairwise differences were statistically significant, although the repeated-measures ANOVA was.

Prior work on SUS validation [8][14][49] established that the average SUS across the spectrum of evaluated systems is 70, and correlated usability levels and grading scales with SUS ranges. Based on this work, the ASL condition's SUS of 71.6 indicates an "OK" level of usability slightly above the threshold for acceptability. Similarly, the Tap condition's SUS of 61.4 falls into the medium range of marginal usability, and the App condition's SUS of 56.3 falls into the low end of marginal usability.

### 4.2 Post-Experiment Survey Results

We asked participants to rate on a five-point Likert scale as to how satisfied they were with their interactions with Alexa when entering movie titles across the three conditions: fingerspelling for ASL, tapping movie title icons on the Echo Show screen for Tap, and entering titles on a touchscreen keyboard for Apps. Figure 5 shows the results. Fingerspelling is preferred, followed by typing on a touchscreen keyboard. In both cases, the majority of participants rated the method favorably. Tapping icons fared less well, with most participants having a neutral or dissatisfied view. Note that due to the nature of Tap-to-Alexa, the movie icons were intermixed with the command icons, and participants had to scroll through the list of command icons to find them.

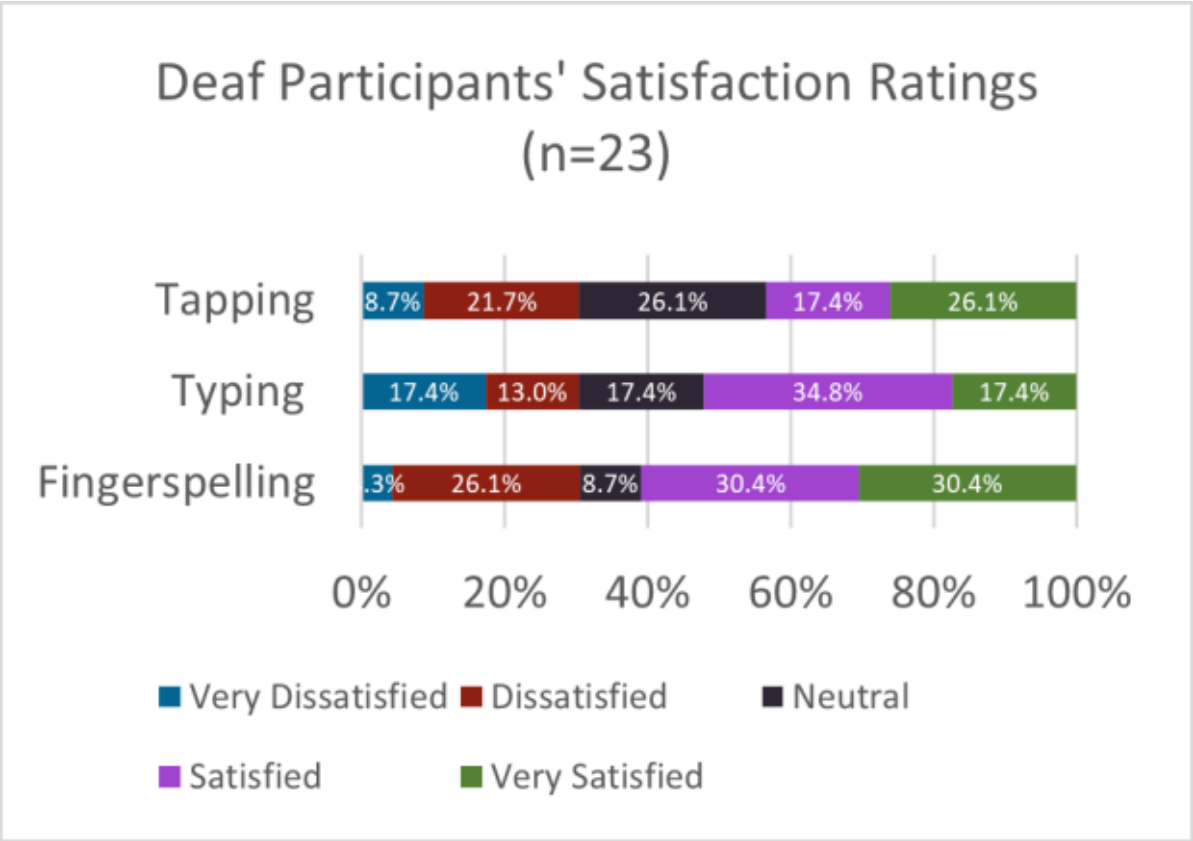

Figure 5: 100% stacked bar graph showing the ratings of the participants' satisfaction for tapping icons, typing using a touchscreen or a keyboard, and fingerspelling movie titles. Fingerspelling had the highest favorability ratings, followed by typing.

We also asked participants directly what input method they could imagine for interacting with an IPA, allowing multiple choices in their responses. The vast majority preferred using ASL to interact with Alexa, as indicated by Figure 6. In the intake survey, those deaf participants also had indicated ASL as their primary mode of communication. More than half of the participants also indicated a willingness to type on a touchscreen or a keyboard if they are unable to utilize ASL. Smart home apps and gestures also were indicated by slightly more than half of the participants. In contrast, fewer than 10% of participants would opt for voice commands when interacting with Alexa.

In Figure 7, we show participants' stated preferences for wake words and methods. The majority prefers eye gaze at the device, combined with either signing a command or waving the hands in an attention-getting gesture. Fewer than half preferred pressing a key, using the Alexa app, or Tap to Alexa as the wake method.

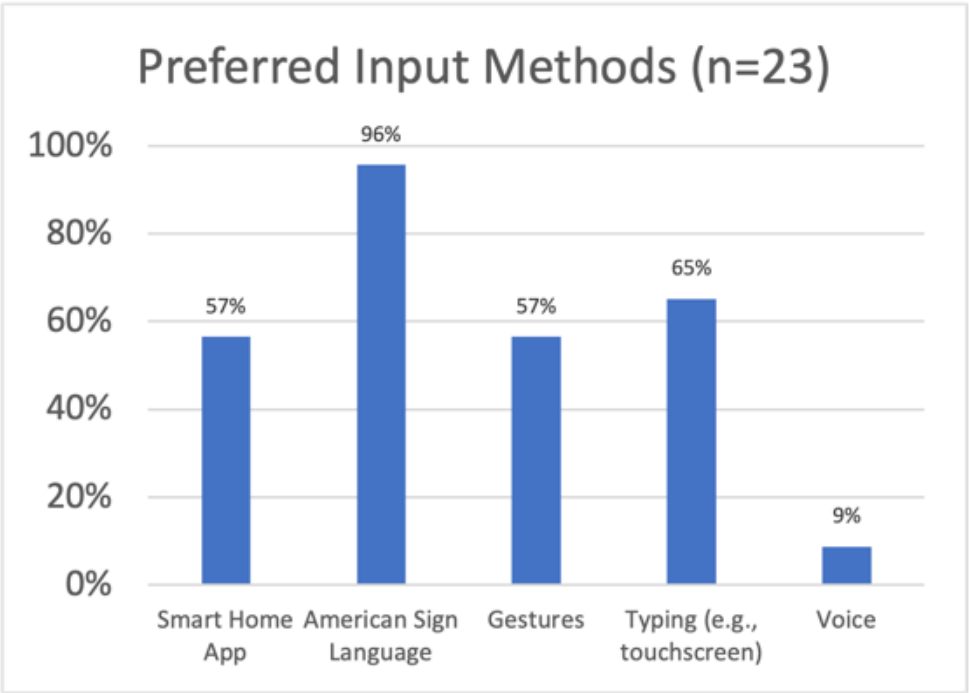

Figure 6: Graph illustrating the preferences of participants in terms of input methods. All but one participant prefers ASL, with typing coming in second.

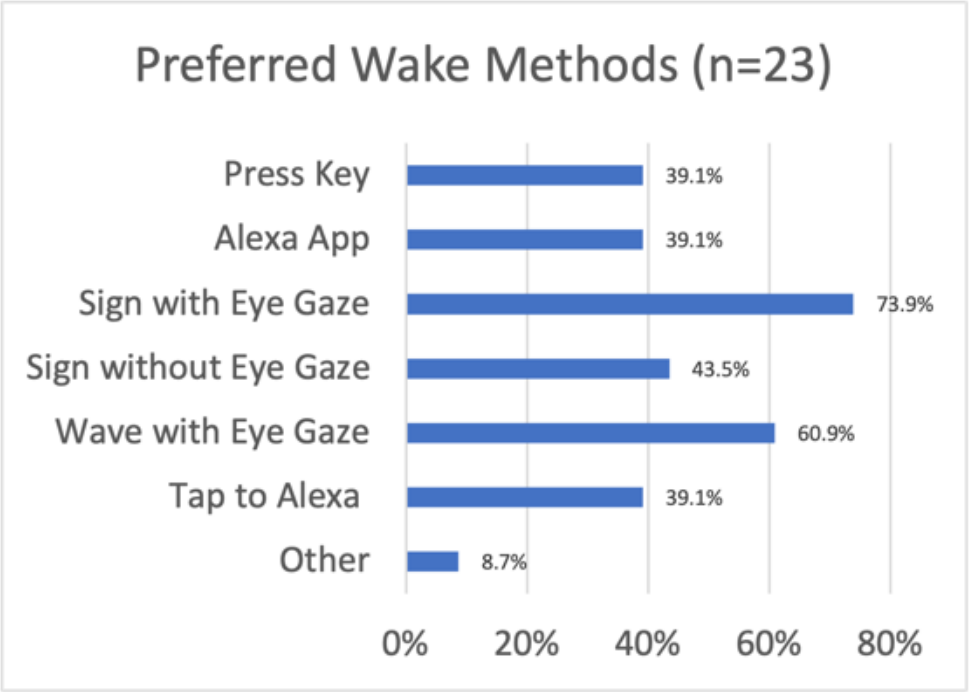

Figure 7: Graph showing preferences for waking Alexa before the start of a command. Signing or waving while looking at the device are by far preferred.

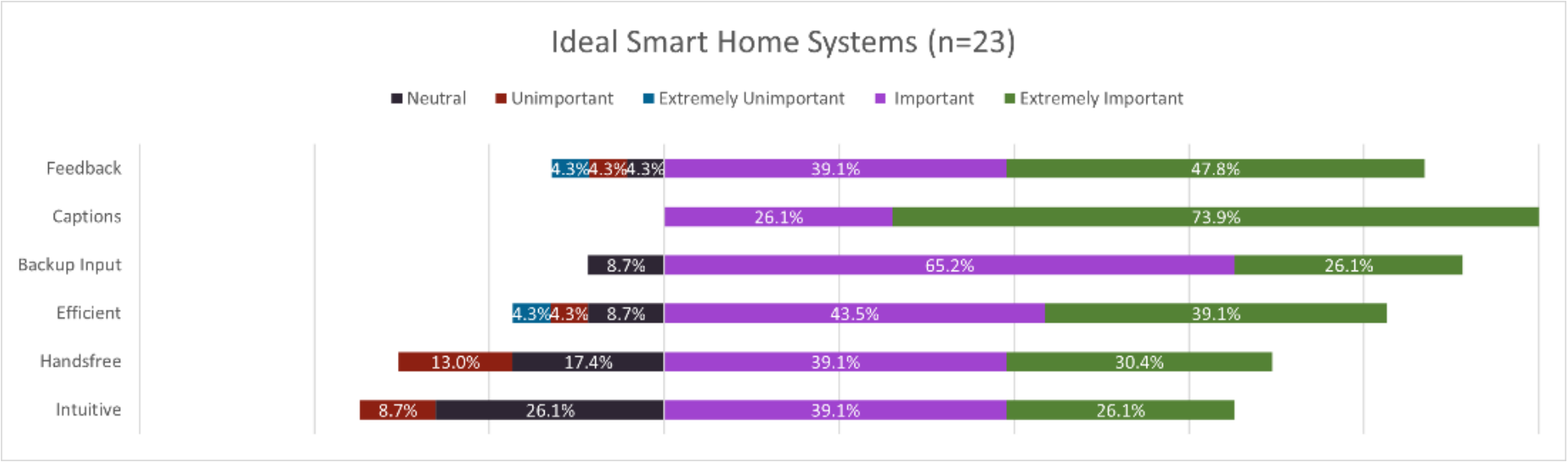

Figure 8: Relative ranking of features needed in an ideal smart home system. Captions were by far the most important one, but the media of each feature still was rated at least as "important."

When asked about features needed in an ideal smart home system, as shown in Figure 8, captions were listed as by far the most important feature, with the requirement that captions include everything spoken, including commands spoken by hearing members within a household. Backup input methods also were important, in case of breakdowns with users' primary input methods. Tactile and visual user feedback, likewise, ranked as important (e.g., being able to see the screen or feel a vibration to know their command was received). Being efficient in issuing commands ranked next, with intuitiveness and hands-free communication options ranking last. However, even those had an overall median rating of “important."

### 4.3 Sign Language Analysis Results

In the annotation and analysis of the signs, we address both the distribution of the signs across participants and specific linguistic phenomena, both of which we expect to affect any future implementations of sign language interaction within IPAs. We annotated a total of 3,645 tokens from the recorded videos.

#### *4.3.1 Distribution and Frequency of Signs*

There was considerable variability across participants with respect to vocabulary size, choice of signs, use of fingerspelling, and gesturing. Figure 9 shows the vocabulary size across participants, broken down into signs, fingerspelled words, and gestures. Note that we had usable video recordings from only 19 out of 23 participants for annotation purposes. Due to the complexities of the experimental setup, we encountered several technical challenges with respect to recording video, which resulted, among other things, in some videos getting cut off prematurely before recording the ASL task.

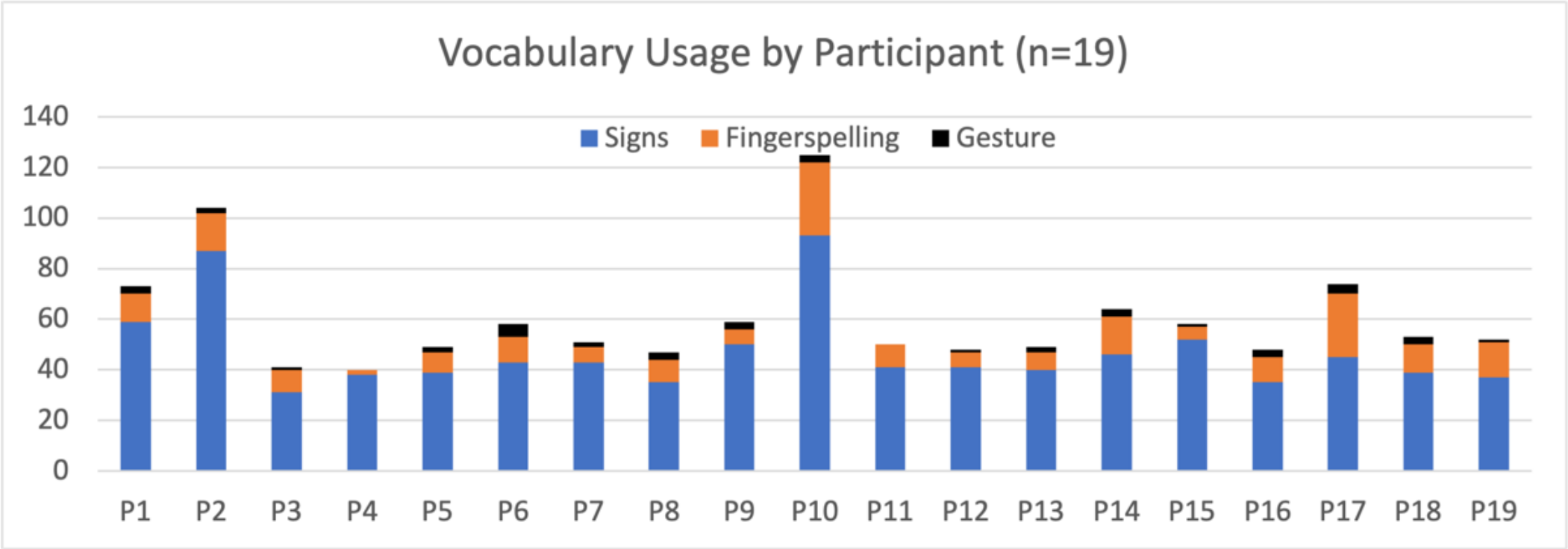

Figure 9: Distribution of vocabulary sizes across participants. The median vocabulary size was 41 signs plus a median of 9 fingerspelled words. A few participants contributed significant outliers to the number of signs in the vocabulary. Additionally, use of finger spelling varied greatly from participant to participant. Almost all participants also used some form of gesturing.

The median number of signs in the vocabulary was 41, while the median number of fingerspelled words was 9. The means were 47+/-17 signs, which were significantly impacted by a few outlier participants, and 10+/-7 fingerspelled words. Participants also employed gesturing. The two most common gesturing were handwaving to get someone’s attention, and extending palms up or outward which, depending on context can denote the end of an action, but also a reaction to something unexpected. Addressing handwaving is important in the context of providing culturally appropriate wake words for IPA commands and is discussed in further detail below.

Across all participants, the total vocabulary size was 246 distinct signs, 93 distinct fingerspelled words, and 18 distinct gestures. 16 signs showed up in multiple variants; for example, the sign for FIRE can be performed by wiggling the fingers while moving the hands straight out, in a circle, or straight up.

Considering that larger vocabularies require proportionately more data in future implementations of sign language recognition systems, we also analyzed each sign as to whether it would be essential in capturing the intent of a command. For example, one third of the participants politely phrased their commands with PLEASE, but recognizing it is immaterial to an IPA understanding and executing a command. Another commonly used sign is THAT which is used as an interjection to express affirmation of a completed event, or to express complete agreement. One common usage scenario occurred after the conclusion of rephrasing a failed command. A third frequently used sign was GO-AHEAD, which prefixes verbs in requests and commands, such as “go ahead and turn on the lights.” In the context of a command-and-control interface with an IPA, the context already makes clear that a command is being issued, and thus recognizing this sign is not essential for in the smart home domain, either. Analysis along these lines revealed that only 117 signs out of the 246 total are essential for the purposes of interacting with an IPA in the smart home domain, given our task lists.

The relative distribution of the most-used signs is shown in the word cloud in Figure 10. Six of these are not essential, as per the discussion in the previous paragraph. Three of these are fingerspelled words; we discuss the nature of fingerspelling in more detail in the next subsection. Another very common sign was indexing – pointing at a referent with the index finger. Its purpose is like the use of “this” and “that” in English, and we discuss it in further detail in the next subsection. Without further data and analysis, it is not clear whether indexing would have to be considered essential for inferring the intent of a command, and how frequently.

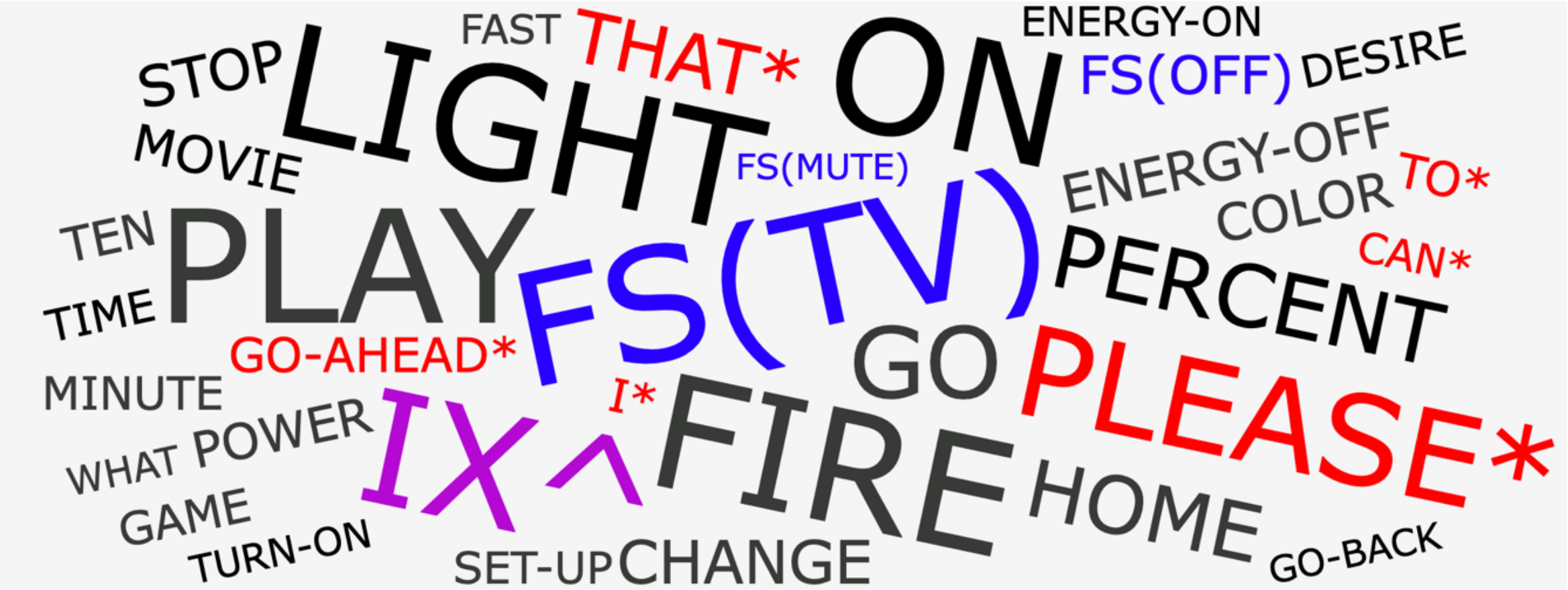

Figure 10: Relative incidence of signs across participants while interacting with Alexa; larger words denote higher incidence. Signs prefixed with FS() in blue are fingerspelled words. IX^ in purple denotes index signs pointing to a specific subject or object referent, analogous to how “this/that” are used in English. Signs with a * suffix written in red are non-key signs; omitting them would not change the meaning of the issued commands.

The distribution of signs exhibited very long tails. 83 signs – more than one-third of the total vocabulary – occurred only once across all participants. Among these, 20 were essential for inferring the intent of commands. Another noteworthy finding is that participants did not always fingerspell the titles of movies, but rather used signs. This accounts for the occurrence of DESIRE and GAME in the list of most common signs in Figure 10, and refers to the movie “Hunger Games.” We will discuss this phenomenon further in the next subsection. Some signs are synonyms in the context of the smart home domain; for example, ON, TURN-ON and ENERGY-ON all function as verbs and have been used interchangeably by the participants for the purposes of turning on lights (but not turning on the TV).

*4.3.2 Observed Linguistic Phenomena*

One of the most-seen linguistic phenomena was fingerspelling, which is often, but not exclusively, used to identify titles and names. The most common item was FS(TV), which represents the fingerspelled word TV and occurred 286 times throughout the 19 participants. However, FS(TV) is also an example of lexicalized fingerspelling; fingerspelling that looks like a sign and may have to be part of the dictionary in a future ASL recognition system. It does not require accurate spelling and often contains omissions, repetitions or movements not typically observed in regular fingerspelling. The lexicalized variants of FS(TV) observed were FS(TV-TV) which shows repetition, FS(FTV) which shows different spelling, and FS(TFTTV) which shows repetition and different spelling. Four instances of FS(T) were also observed which, in context, we can tell are also meant to signal the TV.

The next most common fingerspelled word was FS(OFF), which occurred a total of thirty times, and after that was FS(MUTE) which occurred twenty times. Both also have signs: The sign for OFF occurred 10 times throughout the data, compared with 30 fingerspelled occurrences. For the sign for muting, participants exhibited high variability, including VOICE-OFF, which occurred five times, and gestures, in addition to the fingerspelling of FS(MUTE). FS(SEC), which was used as an alternative to the sign for MOMENT or TIME UNIT, appeared sixteen times. The signs for SECONDS and MINUTES are interchangeable in this context as they both denote small amounts of time; we saw this occur a total of twenty-six times. All together, these examples show clearly that fingerspelling is not just used for names and titles, but also in place of signs. Further examples of participants interchanging both include FS(FORWARD), FS(PAUSE), FS(ON), FS(DIM), all of which have signs.

Conversely, we also noticed participants using signs to refer to movie titles instead of fingerspelling them. For example, a participant may choose to either fingerspell FS(HUNGER GAMES) or sign the concepts for "hunger" and "games." Interpreting this example is context-dependent, because the sign for "hunger" is also the sign for "desire" and "wish," and annotated as DESIRE in the SignBank conventions. With respect to the sign GAME, there was no way of knowing a-priori whether it is plural or singular. Although ASL has ways to indicate GAME as plural, participants did not make use of them to denote the movie title. Hence, this sign, too, required context to determine the appropriate command. Listing the signs for HUNGER/DESIRE/WISH GAME/GAMES in an IPA's vocabulary may not be sufficient to disambiguate if there were a hypothetical video called "Wish Game."

When it comes to the intersection of ASL and English, the rule is not that one word equals one sign, and many signs carry several different meanings. Another example in this experiment, aside from HUNGER/DESIRE/WISH, were the signs for muting the TV. There are many different options, which can denote silencing, crossing out something, and shutting up – all of which were interpreted as "muting." This particular meaning specifically depends on the context in which these signs were used.

Numbers in ASL function in a similar way to lexicalized fingerspelling. We encountered only a few numbers in this study, all between 1-100. These few examples do not capture the full range needed even for a domain-limited application. To cover the full range, numbers would need to be built from fundamental units akin to lexicalized fingerspelling, which spans the signs for 1-9, 10-19, 20, 30, 40, …, 100 (for which there are two common variants), as well as the special conventions for 21, 23, 25 and all duplicated digits 22, 33, 44, …, 99.

Many participants gestured via an attention-wave to the camera, which occurred a total of 60 times. This is a culturally appropriate attention-getting technique within the deaf community. We had not instructed participants to use this attention-getting method, and the Wizard stayed silent when encountering it. However, participants clearly were expecting it to function as an equivalent to using FS(ALEXA) or NS(AX) – a name sign involving the letters A and X in a querying motion – as the wake methods.

The supplemental video accompanying this paper contains footage of participants using ASL to interact with Alexa, illustrating some of the observed phenomena covered in this section.

## 5 DISCUSSION

Our findings indicate that ASL to Alexa demonstrated comparable or better usability compared to English voice input. The SUS for ASL input among deaf users was 71.6 compared to a SUS of 63.7 for spoken English input for hearing users [55]. Both Tap to Alexa and Apps with Alexa input usability were rated worse, with SUS scores falling below 63.7. Most participants expressed challenges in recalling the steps, stating that it would take them a while to get used to Tap to Alexa and Apps with Alexa. Tap to Alexa also forced participants to move closer to the Alexa device, which some found inconvenient. Furthermore, participants found Apps with Alexa to be less intuitive in comparison. While an SUS of 71.6 is better than average, it remains generally low and does not reach the threshold of truly good usability.

The lack of statistical significance in the differences in usability scores suggests that touch-based input options do not clearly lose out over ASL. However, this could also be a result of the limited domain and constraints associated with the task lists. Additionally, the 5-minute window for free play may not have been sufficient to yield conclusive results. However, in the post-experiment survey, participants consistently indicated their preference for ASL input over the text-based alternatives.

### 5.1 Input Methods

A significant number of our participants had no prior experience with Alexa or other IPAs before participating in our study, potentially impacting our results. Several factors could contribute to the relatively low ASL-SUS of 71.6: (1) participants' unfamiliarity with IPAs and a lack of knowledge on how to use them; (2) the limited domain in which participants interacted with Alexa based on the specified task list (Section 3.2.3); and (3) potential language and cultural barriers between the mainstream English-based Alexa system and deaf users. From a hearing perspective, deaf users are considered a cultural and linguistic minority group due to their use of ASL and their connection with the Deaf culture.

Considering that ASL is the primary mode of communication for most participants, there may have been biases that influenced the SUS for ASL. Within the limited time frame of the experiment, this bias could potentially have lowered them, given that participants may initially experience a "struggle" to understand how to use Alexa, as it is a new and unfamiliar interaction for them.

Furthermore, participants are likely more accustomed to app usage, and it is plausible that they might rate apps more favorably during first impressions, due to their familiarity with app interactions. In other words, with increased familiarity with ASL to Alexa, there might have been higher SUS for ASL compared to apps than what we found in this study.

Tap to Alexa presents a unique challenge, as it is a new method of interaction. While touchscreen use is not novel to participants, it presents inconveniences in two ways: having to swipe across screens to search for a command, and physically moving close to the Alexa device. It remains uncertain how these considerations would be affected by familiarity. Further work is needed to answer this question.

Sign language-based interactions are often impractical in everyday situations for deaf users, leading to communication challenges and language barriers [28]. Overall, none of the three input methods tested have SUS that would position them as good enough for everyday usability. Although this study cannot conclusively answer whether ASL is a better input method than the alternatives, it is clear that the usability of current alternatives (apps and Tap) is less than optimal, leaving room for improvement. For some deaf users who can speak, they may prefer to use voice input methods when interacting with IPAs, despite potential challenges posed by their speech limitations (e.g., ASR struggling with deaf speech [24]). In such instances, they may also opt for touchscreen or keyboard input methods as a fallback. Overall, with respect to our research question RQ1 regarding IPA interaction perceptions and preferences, sign language-based interaction holds promise but further work is needed to explore when and how it can supplant other input methods.

In the context of future smart home devices and interaction design, one option to investigate is utilizing Tap to Alexa on Fire tablets for interacting with Alexa, rather than relying on the Echo Show device. This interaction method could enable participants to maintain a distance during interactions instead of moving back and forth between devices. Another possibility is creating a user-friendly, accessible all-in-one app hub instead of requiring users to switch between apps. Although the Alexa app already offers this all-in-one feature, it is complicated and difficult to use. Additionally, wearable devices capable of reading hand gestures [10] would allow Alexa to recognize the wake word, which could be explored. There is rich potential for exploring other types of input methods to improve usability for deaf users.

### 5.2 Sign Language-Based Insights

With respect to research question RQ3, the main takeaways are that: (1) the range of vocabulary and expression among participants in interaction with IPAs is limited on an individual basis; and (2) even combining the vocabulary across participants still yields a limited vocabulary size for the signs that an IPA must recognize in a domain-limited application. Hypothetically, this may make a limited-domain ASL recognition-based IPA feasible, with some important caveats, namely, that fingerspelling, lexicalization of fingerspelling, and the construction of numbers, as well as the observed long tails, may present unexpected challenges. In addition, we found that the meaning of signs can depend on context, which needs to be considered in the bigger picture of a communication session and may pose additional challenges for an ASL-based IPA. In addition, many of the signed commands used indexed signs that point to referents – this is something that needs to be investigated further. It is likely that an IPA must be able to understand what the user is pointing at (e.g., if they are pointing to a specific light in the room that they want to turn off).

Looking at the broader question of the nature of the interaction between deaf users and IPAs, as per RQ2, one of the most crucial takeaways is that IPAs must respect cultural preferences of signers. Many participants used an attention-getting hand wave in hopes to activate the device. Hand-waving in the line of sight of a person, whose attention one is trying to get, is a very common method within the deaf community, as is tapping someone on the shoulder. Using a name sign is also commonly employed, but typically prefaced with a hand-wave to confirm attention. It is possible, even likely, that ASL-based IPAs should not force a specific wake word, and instead support a combination of eye gaze, waving, and the device's name sign (e.g., NS(AX)). Eye gaze itself is an important part of ASL interaction, which functions as a two-way channel of holding attention and receipt of the expressed information. This area will need to be studied further for interacting with IPAs. Furthermore, the results from

Figure 7 are consistent with prior research [41], which found a connection between deaf participants' preference for using ASL over English, and their propensity to interact with signing and waving gestures while maintaining eye gaze.

Additionally, from our results in the post-experiment survey, we see that deaf users are willing to fingerspell when interacting with IPAs and, in fact, prefer it over touch-based input methods, even those that rely on typing. There are a few reasons why this may be; fingerspelling is a common part of ASL and, although it borrows words from English, it allows users to communicate within their native modality of signing. Text-based methods use written English, which can often provide language and cultural barriers to deaf users, which lessens their usability. Secondly, convenience also may play a role. The two touch-based methods differed in their physical set up, and interaction via Tap to Alexa requires users to physically get within touching distance.

As mentioned in distribution and frequency of signs (Section 4.3.1), the median per-person vocabulary size of ASL signs was 41, with an additional median of 9 finger spelled utterances. The medians, however, were greatly affected by the long tail distribution and many of the signs exhibited were only produced once throughout the entire study. The total essential vocabulary size observed in this study was 117 signs across all participants, excluding fingerspelling and the full range of number signs needed. Notably, participants also exhibited a wide variety of fingerspelling and gesture use. The most used gestures throughout the study were an attention-getting wave (which we previously noted may be a more culturally appropriate wake method) and a palm up our outward gesture (which occurred a total of 22 times). The latter typically denotes either the end of an action or a reaction to something unexpected, depending on the context. The third most commonly gesture employed by participants was a filler sign where the signer wiggles their fingers to express that they are thinking, which happened 7 times.

Fingerspelling was often used employed by the participants as a repair strategy throughout the study sessions. However, this strategy is unique to ASL and would not transfer well to many other sign languages. Generally, the use of fingerspelling varies between sign languages, based on the spoken language around them, the cultures associated with the surrounding community, and the history of the signed language. The amount of fingerspelling included in deaf ASL users' everyday language use is reported to be between 12-35 percent, which suggests that ASL users exhibit fingerspelling more than users of other signed languages.

To address signing to Alexa outside an ASL context, mouthing needs to be addressed in future work. It is often seen along or in lieu of a fingerspelled word, both in ASL and in other signed languages. Much like ASL, for example, British Sign Language (BSL) users' mouth in combination with a sign to denote which of the several meanings the sign can represent is being used currently. Mouthing and lip-reading are relevant to fingerspelling as well, not always but often, ASL signers will mouth the word they are fingerspelling [38].

### 5.3 Alexa Response Challenges

In addressing RQ2, interacting with IPAs is a complex process that involves both input and output. We need to consider not only the commands and information provided by deaf users (input) but also the responses and feedback provided by the IPAs (output). This consideration is pivotal in evaluating how deaf users interact with the IPA technology. In terms of the output aspect, about 72% of the participants perceived captions as an extremely important feature of ideal smart home systems. Captions are vital for providing visual feedback and facilitating information access. Note that there is a consistent lack of caption availability in the current IPA technology. Researchers in the field, especially those who do not engage with deaf people, may not be fully aware of the need for captions [15]. Despite the presence of captions on the Amazon Echo Show, deaf participants may frequently miss them or be unable to understand what Alexa is saying. This is primarily due to the frequent instances where deaf participants have trouble reading the captions or find them to be inadequately legible, making it challenging to understand what Alexa is trying to say according to prior work [52]. Our study assumed a priori that the existing Alexa captions were usable. However, considering the importance attributed by participants, additional assessments are necessary to determine the effectiveness and user-friendliness of the captions provided by IPAs. It may also be necessary to consider alternatives to captions.

## 6 LIMITATIONS AND FUTURE WORK

### 6.1 Limitations

Many of the signs used show potential for interpreter error, for example the sign used for one of the Fire TV functions of playing a movie, The Hunger Games (annotated DESIRE GAME). The sign for “hunger” is the same sign as for “desire,” and its correct interpretation depends on the Wizard’s understanding of the context. Not all fingerspelling is crisp and clear, and participants sometimes produced fingerspelled words with partially produced or omitted letters. While the Wizard had been trained on the limited domain, and thus had sufficient information to infer the correct words from participants’ unclear signing, there will always be potential for human error that differs from how a hypothetical ASL recognition system may behave.

Another limitation is that we required participants to use a wake word, which is currently an English-based concept. As mentioned in Section 5.2, this is not in line with sign language communication expectations. As a result, there have been periods of user frustration, which may have affected usability perceptions. The sign language to spoken English chain of transmission for commands in the Wizard-of-Oz setup also is a potential limiting factor. Due to the interpreting, there was additional lag time between the ASL input of the participant's command and Alexa's response and corresponding processing, compared to voice-only interaction. The need for audio from the Wizard to issue commands also resulted in problems and frustration that would not exist in a hypothetical ASL-based IPA. For example, one participant experiment was interrupted by Alexa misunderstanding a command and proceeding to play music loudly enough that the Wizard from the other room could not speak over it to override the command. All these issues may have impacted usability scores of the ASL to Alexa method.

### 6.2 Future Work

Future work should explore the studies of culturally appropriate wake-methods in more depth. Researchers might consider investigating the effectiveness of attention-getting gestures, such as waving, in conjunction with eye gaze as an alternative wake-up method that aligns more with the cultural preferences of deaf users. The ASL capabilities in IPAs could be expanded to include interactions in a kitchen setting, for instance, instructing an IPA to find a pasta recipe and follow the accompanying video instructions. Additionally, exploring ASL interactions in other day-to-day settings, such as in the car, could be valuable. As smart home technology is becoming more prevalent in home environments and daily life, allowing deaf interactions with IPAs is an initial step towards ensuring equal access for deaf users in smart home environments.

Our findings (Section 4.3.2) regarding the vocabulary size exhibited by participants in this limited-domain application indicate that IPA systems may be able to interact with deaf users utilizing constrained predetermined word banks. This could be accomplished by supplementing these systems with separate components designed for fingerspelling and number recognition, which incorporate the unique conventions associated with these aspects of sign language. The components designed for fingerspelling and number recognition would require a more nuanced focus on fingerspelling and numbers within the experimental design itself to obtain more data. At this point in our research, we have noticed high variability in fingerspelling use amongst participants. To get a more accurate data pool relating to fingerspelling conventions, we suggest making fingerspelling use a requirement to the participants either by asking them to communicate concepts that have no set sign for them, or by requiring fingerspelling as the method of communication for whatever they are asked to communicate.

Given the prevalent use of IPAs in homes, it is important for future research to replicate similar conditions. While the study was stimulated in a living room environment, it is worth noting that most households include more than one room (e.g., kitchen, living room and bedroom). Researchers might consider exploring the common rooms and activities that people typically engage in at home. Additionally, they should conduct usability testing that involves transitioning between different rooms within a household to gauge its impact. One specific domain that we recommend tackling is the kitchen environment. This environment presents a distinct challenge as communication with an IPA will likely vary when users have dirty hands cooking in the kitchen. Furthermore, there is a need to explore other everyday applications beyond home environments, such as in the car.

Finally, this work confirms the importance of interdisciplinary collaboration [13], requiring expertise in technology, accessibility, interpreting, and ASL linguistics. We drew on interdisciplinary expertise during the design of the Alexa setup, addressing interpreting challenges, respecting the cultural factors that need to go into UI design, and implementing a standardized annotation system.

## 7 CONCLUSION

Although this study did not confirm that ASL is preferred over touch-based input methods, it did substantiate that ASL needs to be considered as an option for IPAs, pending further research. The SUS for ASL to Alexa is within the average range, bordering on acceptability; however, it still surpasses the SUS for similar English voice systems. Specifically, this study found the median ASL vocabulary size of participants to be 41, and only 117 out of the 246 of the signs produced by participants throughout their portion were deemed essential for IPA system's understanding of their commands. This finding suggests that limited-domain interaction with automatic ASL recognition in IPAs may be feasible, although several important linguistic phenomena would need to be addressed. These include fingerspelling, interchanging fingerspelling and signing, numbers, indexing, eye gaze, and gesturing. Additionally, the English-based wake-word method currently employed by personal assistant technologies should be revisited – more culturally appropriate wake methods, such as an attention-getting wave, should be considered. Lastly, the study findings on the nuances of ASL usage with IPAs demonstrate the value of tackling the problem from an interdisciplinary standpoint, as also suggested in [13].

## ACKNOWLEDGMENTS

The contents of this paper were developed under a grant from the National Institute on Disability, Independent Living, and Rehabilitation Research (NIDILRR grant number 90REGE0013). NIDILRR is a Center within the Administration for Community Living (ACL), Department of Health and Human Services (HHS). The contents of this site do not necessarily represent the policy of NIDILRR, ACL, HHS, and you should not assume endorsement by the Federal Government. Norman Williams supported the technical setup of the experiment. Vasu Kushalnagar assisted with the audio setup. Kathleen Rehagen and Haili Balderson provided support for the Wizard role. James Waller provided feedback on the statistical analysis.

# A APPENDICES

## A.1 Task List A

1. Turn on the lights.
2. Turn on the Fire TV.
3. Play The Rings of Power on the Fire TV.
4. Mute the Fire TV.
5. Set a timer for 2 minutes.

6. Play [a movie or show from the list below] on the Fire TV.
7. Dim the lights 75%.
8. Pause the Fire TV.
9. Resume the Fire TV.
10. Rewind 30 seconds on the Fire TV.
11. Fast forward 10 seconds on the Fire TV.
12. Go home on the Fire TV.
13. Turn up the lights. (Give a percentage.)
14. Change the light color. (Give a color.)
15. Turn off the lights.
16. Turn off the Fire TV.

### A.2 Task List B

1. Turn on the Fire TV.
2. Turn on the lights.
3. Play the Rings of Power on the Fire TV.
4. Mute the Fire TV.
5. Dim the lights 50%.
6. Set a timer for one minute.
7. Play [a movie or show from the list below] on the Fire TV.
8. Fast forward 10 minutes.
9. Pause the Fire TV.
10. Resume the Fire TV.
11. Rewind 3 minutes on the Fire TV.
12. Change the light color. (Give a color.)
13. Go home on the Fire TV.
14. Turn up the lights. (Give a percentage.)
15. Turn off the lights.
16. Turn off the Fire TV.

### A.3 Task List C

1. Turn on the lights.
2. Turn on the Fire TV.
3. Play The Rings of Power on the Fire TV.
4. Mute the Fire TV.
5. Play [a movie or show from the list below] on the Fire TV.
6. Dim the lights 25%.
7. Set a timer for 3 minutes.
8. Rewind 1 minute on the Fire TV.
9. Fast forward 5 minutes on the Fire TV.
10. Pause the Fire TV.

11. Resume the Fire TV.
12. Turn up the lights. (Give a percentage.)
13. Change the light color. (Give a color.)
14. Go home on the Fire TV.
15. Turn off the lights.
16. Turn off the Fire TV.

### A.4 MOVIE & SHOW LIST [Amazon Prime only]:

The Hunger Games
The Legend of Vox Machina
The Report

### A.5 Detailed Experimental Procedures

The limited-domain smart home environment setup included an Amazon Echo Show device configured to display Alexa's responses as captions. Tap-to-Alexa [3] was also enabled, allowing users to interact with Alexa via the touchscreen using a preconfigured set of commands instead of voice input. We provided an Amazon Fire TV connected to the Echo Show device, facilitating video playback. Additionally, two Philips Hue multicolor lights were provided, linked to the Echo Show device and configured to blink and change colors in response to user-set alarms or timers (see also Figure 1 in Section 3.3.1).

For participants to interact with the system through smart home apps, we provided a 9th-generation Apple iPad. The iPad had preinstalled Alexa, Fire TV [4], and Philips Hue [45] apps, all connected to this Internet of Things environment. To record participant behaviors and actions during their interactions with Alexa, an HD camera was mounted on the top of the Echo Show device, while a rear camera was placed within the stimulated living room environment. We also provided a remote for additional control of the Fire TV if the participants preferred to use it, for example, in cases where the Fire TV remote app was not functioning.

The Wizard-of-Oz design had critical requirements: (1) clear visibility of the participant's signing for an off-screen American Sign Language (ASL) interpreter (referred to as the "Wizard"), (2) maintaining a clear audio connection between the Wizard and the Echo Show device, and (3) ensuring that the participant remained unaware of the presence of the Wizard. Note that the participants were unaware that the camera was connected to a laptop, rather than, as they believed, the Alexa system itself. The intention behind this design decision was to avoid providing participants with more information than was necessary for the experiment.

To fulfil the initial two criteria, we configured two MacBook Air laptops. One was placed in the participant's room, referred to as "Dorothy," and the other was in a separate room designated as the "Wizard" room. These laptops were interconnected via hardwired Ethernet to ensure unimpeded audio and video transmission. The Dorothy laptop was equipped with the above-mentioned HD webcam and an EarFun UBOOM 28W speaker, which replaced the built-in laptop speakers for clearer audio output. This adjustment was necessary as the built-in speakers were not sufficiently clear to consistently trigger Alexa commands. The Wizard laptop was equipped with a Blue Yeti microphone to ensure maximal clarity for the ASL interpreter's voice during the ASL-to-English spoken translation.

To fulfil the third criterion, the screen of the Dorothy laptop screen was placed behind the Fire TV and angled away. The researcher engaging with the participant refrained from any interaction with that laptop throughout the session to prevent its visibility. By angling away the screen, we prevented the participants from becoming aware of the FaceTime link, or potentially from noticing the laptop in the first place. We did this to remove a potential source of bias induced by participants becoming aware of a system separate from Alexa.

The Wizard had the responsibility for controlling both laptops, using VNC Viewer app [47] to remotely control the Dorothy laptop. The Wizard established a FaceTime link [7] between both laptops to observe the participant's signing through the HD webcam at the optimal framerate and resolution. Additionally, the Wizard employed the Photo Booth app [6] on the Dorothy laptop via VNC to locally record the webcam video at the native frame rate. This setup allowed the Wizard to observe, record, and interpret for the participants without their knowledge, all while being able to hear Alexa's audio responses. Additionally, if the participant did not sign the wake word, the Wizard refrained from saying the wake word accordingly (Section 3.3.2). In instances where researchers needed to communicate clandestinely for troubleshooting purposes, they utilized the Discord app on their mobile phones.